\documentclass[sigconf]{acmart}
\usepackage{multirow}
\usepackage{graphicx}
\usepackage{caption}
\usepackage{subcaption}
\usepackage{stfloats}
\usepackage{enumitem}
\usepackage{hyperref}

\AtBeginDocument{%
  }

\setcopyright{acmcopyright}
\copyrightyear{2022} 
\acmYear{2022} 
\setcopyright{acmcopyright}\acmConference[CIKM '22]{Proceedings of the 31st ACM International Conference on Information and Knowledge Management}{October 17--21, 2022}{Atlanta, GA, USA}
\acmBooktitle{Proceedings of the 31st ACM International Conference on Information and Knowledge Management (CIKM '22), October 17--21, 2022, Atlanta, GA, USA}
\acmPrice{15.00}
\acmDOI{10.1145/3511808.3557154}
\acmISBN{978-1-4503-9236-5/22/10}

\begin{document}
\bibliographystyle{ACM-Reference-Format}

\title{Scenario-Adaptive and Self-Supervised Model for Multi-Scenario Personalized Recommendation}


\author{Yuanliang Zhang}
\authornote{Both authors contributed equally to this research.}
\affiliation{%
  \institution{Alibaba Group}
  \city{Hangzhou}
  \country{China}
}
\email{kubert.zyl@alibaba-inc.com}

\author{Xiaofeng Wang}
\authornotemark[1]
\affiliation{%
  \institution{Alibaba Group}
  \city{Hangzhou}
  \country{China}
}
\email{aron.wxf@alibaba-inc.com}

\author{Jinxin Hu}
\affiliation{%
  \institution{Alibaba Group}
  \city{Hangzhou}
  \country{China}
}
\email{jinxin.hjx@alibaba-inc.com}

\author{Ke Gao}
\affiliation{%
  \institution{Alibaba Group}
  \city{Hangzhou}
  \country{China}
}
\email{gaoke.gao@alibaba-inc.com}

\author{Chenyi Lei}
\affiliation{%
  \institution{Alibaba Group}
  \city{Hangzhou}
  \country{China}
}
\email{chenyi.lcy@alibaba-inc.com}

\author{Fei Fang}
\affiliation{%
  \institution{Alibaba Group}
  \city{Hangzhou}
  \country{China}
}
\email{mingyi.ff@alibaba-inc.com}

\renewcommand{\shortauthors}{Zhang et al.}

\begin{abstract}
Multi-scenario recommendation is dedicated to retrieve relevant items for users in multiple scenarios, which is ubiquitous in industrial recommendation systems. These scenarios enjoy portions of overlaps in users and items, while the distribution of different scenarios is different. The key point of multi-scenario modeling is to efficiently maximize the use of whole-scenario information and granularly generate adaptive representations both for users and items among multiple scenarios. we summarize three practical challenges which are not well solved for multi-scenario modeling: (1) Lacking of fine-grained and decoupled information transfer controls among multiple scenarios. (2) Insufficient exploitation of entire space samples. (3) Item's multi-scenario representation disentanglement problem. In this paper, we propose a \textbf{S}cenario-\textbf{A}daptive and \textbf{S}elf-\textbf{S}upervised (\textbf{SASS}) model to solve the three challenges mentioned above. Specifically, we design a \textbf{M}ulti-\textbf{L}ayer \textbf{S}cenario \textbf{A}daptive \textbf{T}ransfer (\textbf{ML-SAT}) module with scenario-adaptive gate units to select and fuse effective transfer information from whole scenario to individual scenario in a quite fine-grained and decoupled way. To sufficiently exploit the power of entire space samples, a two-stage training process including pre-training and fine-tune is introduced. The pre-training stage is based on a scenario-supervised contrastive learning task with the training samples drawn from labeled and unlabeled data spaces. The model is created symmetrically both in user side and item side, so that we can get distinguishing representations of items in different scenarios. Extensive experimental results on public and industrial datasets demonstrate the superiority of the SASS model over state-of-the-art methods. This model also achieves more than 8.0\% improvement on Average Watching Time Per User in online A/B tests. SASS has been successfully deployed on multi-scenario short video recommendation platform of Taobao in Alibaba.
\end{abstract}




\maketitle

\section{Introduction}

\begin{figure}[!tbp]
  \setlength{\abovecaptionskip}{3pt}
  \setlength{\belowcaptionskip}{1pt}
  \centering
  \includegraphics[width=0.9\linewidth]{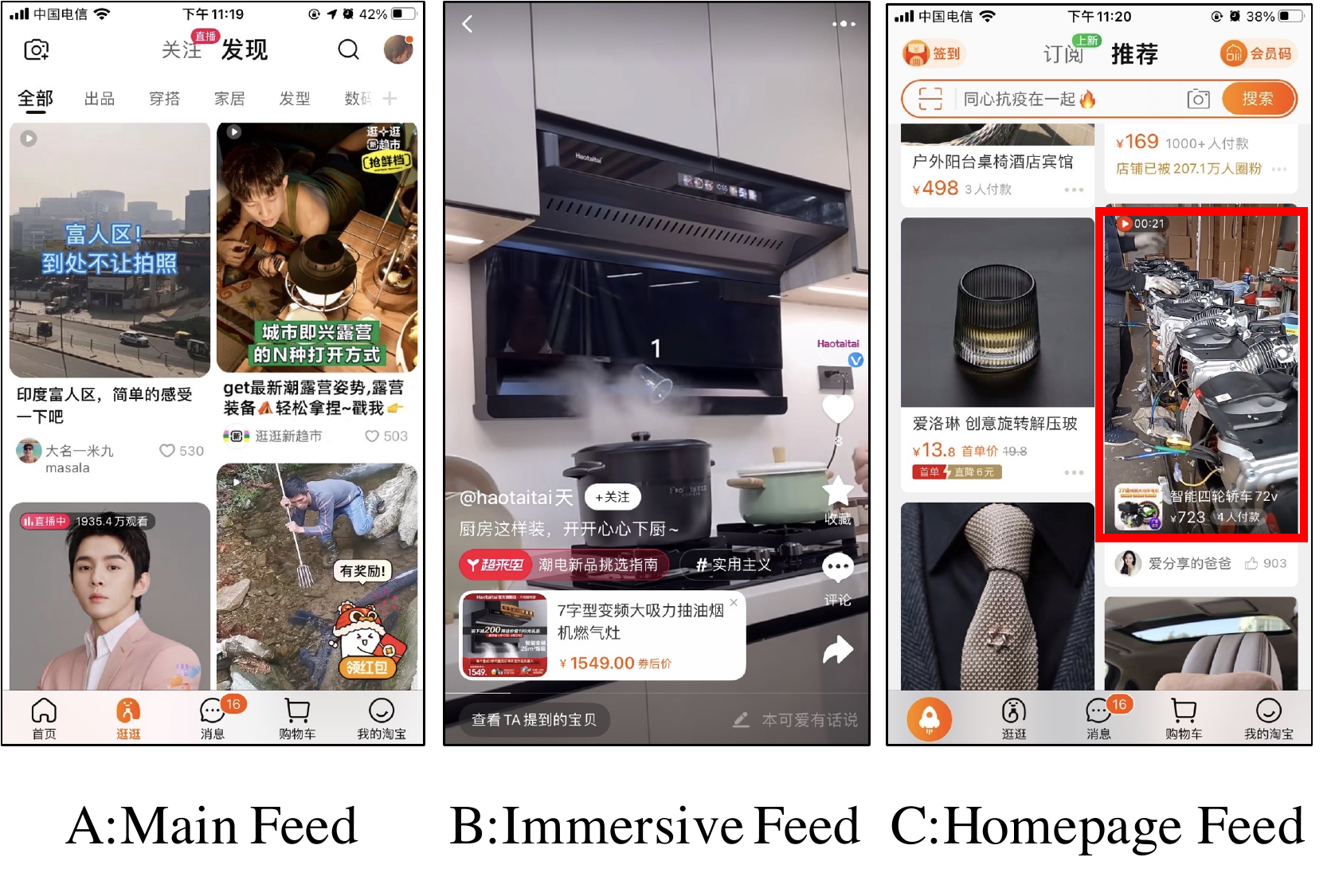}
  \caption{Short video recommendation scenarios in Taobao.}
  \label{fig:scenario_examples}
\end{figure}

In recent years, the \emph{Multi-Scenario Personalized Recommendation Problem}\cite{sheng2021one,li2020improving,chen2020scenario,shen2021sar,zhang2022leaving}, which focuses on retrieving relevant candidates in multiple scenarios, is well-known and ubiquitous in most industrial recommendation systems, such as Taobao in Alibaba, Amazon, TikTok etc. The \emph{scenario} can be treated as a specific recommendation domain of users and items. As shown in Figure \ref{fig:scenario_examples}, there are diverse scenarios in short video recommendation platform of Taobao. From the user perspective, a user may access some of the scenarios and watch different videos. Form the video perspective, a short video may be pushed to different users in different scenarios. There are some common users and videos among different scenarios, making it reasonable and beneficial to share information for model learning. However, each scenario will also have its own unique users and videos. Besides, The behaviors are diverse for the same user in different scenarios, and the exposure abilities of the same video in different scenarios are also distinguishing. Therefore, it is challenging to model the commonalities and distinctions of different scenarios when solving the multi-scenario problem.

\begin{figure*}[tb]
  \setlength{\abovecaptionskip}{1pt}
  \setlength{\belowcaptionskip}{3pt}
  \centering
  \includegraphics[width=\linewidth]{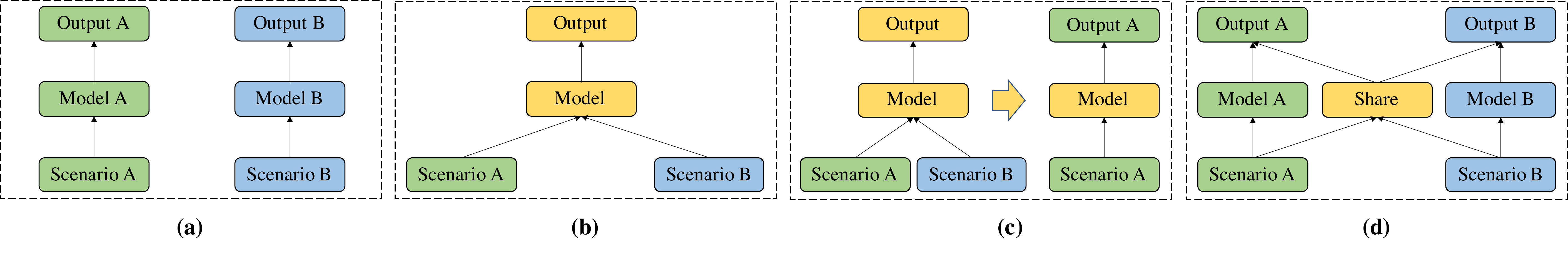}
  \caption{Overview of existing strategies for multi-scenario modeling. (a) Training a single model for each scenario with its own data. (b) Training a common model using multi-scenario data. (c) Pre-training a common model with multi-scenario data and fine-tuning with single scenario data for each scenario. (d) Building a unified model with the framework of multi-task learning.}
  \label{fig:traditional_methods}
  \vspace{-10pt}
\end{figure*}

Various types of strategies have been proposed to tackle the multi-scenario problem: (1) Training a separate model for each scenario (depicted as Figure \ref{fig:traditional_methods}(a)). The shared information among multiple scenarios is neglected in this strategies. It is challenging for new and minor scenarios with limited data to learn proper models. Besides, developing a separate model for each scenario will consume tremendous resources. (2) Training one common model with mixture of samples coming from multiple scenarios (depicted as Figure \ref{fig:traditional_methods}(b)). The operation of sample mixture destroys the original data distribution, making it hard for one model to make appropriate predictions for each scenario. Moreover, minor scenario may be dominated by major scenario. (3) Training a model with whole-scenario samples and fine-tuning scenario-specific models with samples from corresponding scenario, depicted as Figure \ref{fig:traditional_methods}(c). This method can enjoy whole-scenario data and get scenario-specific results. However, little attentions are paid to model the correlations among multiple scenarios. (4) Building a unified framework based on multi-task learning and learning commonalities and correlations among multiple scenarios\cite{shen2021sar,li2020improving,zhang2022leaving} (depicted as Figure \ref{fig:traditional_methods}(d)), which has become a mainstream strategy to solve multi-scenario problem. However, previous works overlook three crucial challenges:

\begin{itemize} [topsep=2pt, leftmargin=8pt]
    \item \textbf{Lacking of fine-grained and decoupled information transfer controls}. The key point of learning scenario correlations is to model the information transfer among scenarios. Existing works conduct information transfer in quite implicit and coarse-grained ways, such as parameter factorization\cite{sheng2021one}, mixture-of-experts mechanisms\cite{shen2021sar,li2020improving} and dynamic weights network\cite{zhang2022leaving}. it is hard to determine the precise magnitude of transferred information from one scenario to another.
    \item \textbf{Insufficient exploitation of entire space samples}. Most existing multi-scenario models are trained only with labeled data, making huge amount of unlabeled data (users or items without interactive behaviors in a specific scenario) under no use. These unlabeled data are of great potential for reflecting scenario characteristics. It is also challenging to model some spare or new scenarios with little labeled data. Although ZEUS\cite{gu2021self} has made some attempts to model unlabeled data through a self-supervised \emph{Next-Query Prediction} task, it only focuses on query-side modeling. Besides, there are little unions between the self-supervised task and multi-scenario modeling in ZEUS.
    \item \textbf{Item's multi-scenario representation disentanglement problem}. From the perspective of item side, an item may have distinguishing characteristics and behaviors in different scenarios. It is quite suitable to generate distinguishing representations for items in different scenarios. To the best of our knowledge, previous methods mainly focus on scenario-aware intention modeling from user perspective, with little concerns on the item side. 
\end{itemize}

In order to tackle the challenges mentioned above, we propose the Scenario-Adaptive and Self-Supervised (SASS) model and demonstrate its rationality and effectiveness. We design a Multi-Layer Scenario Adaptive Transfer (ML-SAT) module to learn user and item representations in scenario-specific scope and whole-scenario scope. In ML-SAT a scenario-adaptive gate unit with explicit gate controls is proposed to regulate and facilitate the fine-grained information transfer from global shared network to scenario-specific network. To adequately exploit the entire space samples, we introduce a two-stage training process (pre-training and fine-tune). In pre-training stage, we build a scenario-supervised contrastive learning task with the training samples drawn from labeled and unlabeled data spaces. In fine-tune stage, the user-item similarity matching objective is achieved with a scenario-specific and a global-auxiliary matching task. The model architecture of SASS is created symmetrically both in user side and item side, so that we can get distinguishing representations for items in different scenarios. We creatively combine the multi-scenario problem and self-supervised contrastive learning problem in a unified paradigm and tackle the challenges of multi-scenario modeling mentioned above.

The main contributions of this work are summarized as follows:

\begin{itemize} [topsep=2pt, leftmargin=8pt]
    \item We propose an effective Scenario-Adaptive and Self-Supervised (SASS) framework to solve multi-scenario problem. In SASS, we design a Multi-Layer Scenario Adaptive Transfer module with a scenario-adaptive gate unit to regulate and facilitate the fine-grained information transfer from global shared network to scenario-specific network.
    
    \item We introduce a two-stage training process to strengthen the exploitation of entire space samples, especially unlabeled data in corresponding scenario. We design a novel scenario-supervised contrastive learning task and closely correlate the scenario-supervised task with multi-scenario problem.
    
    \item The proposed model structure is symmetrically designed in item side, so that we can generate distinguishing representations for items in different scenarios.
    
    \item Extensive experimental results on public and industrial datasets demonstrate the superiority of the SASS model over state-of-the-art methods. SASS has also been successfully deployed on multi-scenario short video recommendation platform of Taobao in Alibaba and achieved more than 8.0\% improvement on Average Watching Time Per User in online A/B tests. We believe the strategies proposed in SASS are universally applicable in most multi-scenario recommendation systems.
\end{itemize}

\section{Related Work}

\subsection{Multi-Scenario Recommendation Models}
As mentioned above, the mainstream strategies of tackling the multi-scenario problem is to create a unified framework to model all scenarios simultaneously. Thus, we mainly survey works related to this paradigm. Specifically, HMoE\cite{li2020improving} utilizes multi-gate mixture-of-experts\cite{ma2018modeling} to implicitly model commonalities and distinctions among multiple scenarios. SAML\cite{chen2020scenario} distinguishes user behaviors in different scenarios with scenario-specific attention mechanisms and proposes a scenario-mutual unit to learn differences and similarities between scenarios. ICAN\cite{xie2020internal} treats each data channel as a scenario and designs a scenario-aware contextual attention layer to generate distinguishing user representations in different scenarios. SAR-Net\cite{shen2021sar} proposes a unified multi-scenario architecture and introduces two scenario-aware attention modules to extract scenario-specific features in user side. After that, SAR-Net implements implicit scenario information transfer with the gate fusion of scenario-specific experts and scenario-shared experts. STAR\cite{sheng2021one} designs a star topology framework, with one centered network to maintain whole-scenario commonalities and a set of domain-specific networks to distinguish scenario distinctions. The combination strategy of element-wise product of layer weights is treated as the information transfer mechanism from whole scenarios to individual scenario. M2M\cite{zhang2022leaving} pays attention to advertiser modeling in multiple scenarios and proposes a dynamic weights meta unit to model inter-scenario correlations. The methods mentioned above learn scenario information transfer in quite implicit ways, making it hard to determine the precise impacts among multiple scenarios. Besides, these models are only trained with labeled data, without making full use of the entire space samples. Although ZEUS\cite{gu2021self} tackles the feedback loop problem with a \emph{Next-Query Prediction} self-supervised manner on unlabeled data, the pre-training task is mainly based on user spontaneous query sequences. Moreover, all of the previous works overlook the problem of generating distinguishing representations of items in different scenarios. 

\subsection{Self-Supervised Learning in Recommendation System}

Self-supervised learning which focuses on learning feature embeddings and initial network weights on unlabeled data, is widely utilized in areas of Compute Vision\cite{chen2020simple,chen2021exploring,pathak2016context} and Natural Language Processing\cite{devlin2018bert,gururangan2020don}. Many efforts based on self-supervised learning have also been made in the area of Recommendation Systems. \cite{ma2020disentangled} proposes a sequence-to-sequence self-supervised training strategy for sequential recommendation, while \cite{sun2019bert4rec} models user behavior sequences by predicting the random masked items in the sequence with bidirectional self-attention. SGL\cite{wu2021self} proposes a self-supervised task on user-item graph with various operators to generate different graph views. To tackle the label sparsity problem, \cite{yao2021self} introduces the self-supervised strategy based on the contrastive learning on the item side. CLRec\cite{zhou2021contrastive} also utilizes contrastive learning to reduce the exposure bias in deep candidate generation. S$^{3}$-Rec\cite{zhou2020s3} models sequence recommendation by building four auxiliary self-supervised objectives with the mutual information maximization principle. Moreover, ZEUS\cite{gu2021self} proposes a \emph{Next-Query Prediction} self-supervised task on user's query sequences for spontaneous search Learning-To-Rank tasks.
 

\begin{figure*}[tb]
  \setlength{\abovecaptionskip}{0pt}
  \setlength{\belowcaptionskip}{1pt}
  \centering
  \includegraphics[width=\linewidth]{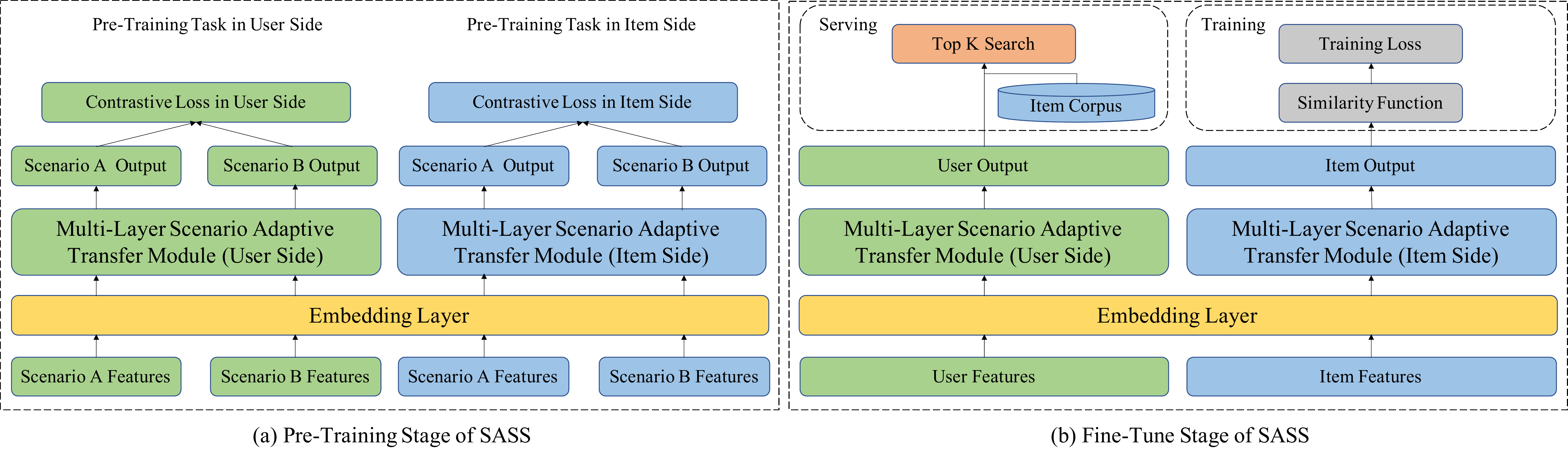}
  \caption{Overall architecture of SASS model. (a) Pre-training stage of SASS. (b) Fine-tune stage of SASS}
  \vspace{-10pt}
  \label{fig:architecture}
\end{figure*}

\section{Problem Formulation}

We propose a unified framework to solve multi-scenario problem, which is definitely applicable to serve both matching and ranking tasks in recommendation system. In this paper, we describe and demonstrate our method in matching task. Few modifications can be made when it is served for ranking tasks, which we remain for future research.



\subsection{Multi-Scenario Matching Problem}

A \textbf{Scenario} can be treated as a specific recommendation domain, denoted as $D_{s}$ in this paper. Given a set of scenarios $D = {\{D_s\}^{|D|}_{s=1}}$, which share a common feature space $\mathcal{F}$ and label space $\mathcal{Y}$. For scenario $D_{s}$, the labeled training data are drawn from a domain-specific distribution $\mathcal{P}_s$ over $\mathcal{F} \times \mathcal{Y}$. Although there may be many common users and items between two scenarios, the distribution $\mathcal{P}_s$ are quite distinguishing in different scenarios. The \textbf{Multi-Scenario Matching Problem} can be formulated as follows:
\begin{equation}
    \setlength{\abovedisplayskip}{3pt}
    \setlength{\belowdisplayskip}{3pt}
    \widetilde{V}_s = max_{1 \le k \le K}(sim(e_u^s, e_v^s)), v \in V
\end{equation}
where $V$ denotes the large-scale candidate item set. $e_u^s$, $e_v^s$ is the user and item representation vectors in scenario $D_{s}$, $sim(e_u^s, e_v^s)$ is the relevance score between user $u$ and item $v$, and $\widetilde{V}_s$ is the final top $K$ matching results for $D_{s}$.

The multi-scenario problem can be treated as a specific case of \textbf{multi-domain learning problem} with three important characteristics: (a) All scenarios have the same user/item type and enjoy a common feature schema; (b) All scenarios share the same learning objective (multiple inputs and one objective); (c) The key points of multi-scenario learning focus on the information sharing and transfer among scenarios, so that all scenarios' performances can be improved simultaneously. It should be emphasized that the multi-scenario problem is quite different from cross domain problem\cite{hu2018conet,li2020ddtcdr,ouyang2020minet,xie2021contrastive} (mainly focusing on improving target domain's performance), multi-task learning problem\cite{ma2018modeling,misra2016cross,tang2020progressive} (one input and multiple objectives) and heterogeneous multi-domain task\cite{hao2021adversarial} with different item types.


\section{Proposed Method}


In this section, we describe the proposed Scenario-Adaptive and Self-Supervised Model (SASS). SASS has two stages, Pre-Training Stage and Fine-Tune Stage:
\begin{itemize} [topsep=2pt, leftmargin=8pt, partopsep=2pt]
    \item \textbf{Pre-Training Stage}, as shown in Figure \ref{fig:architecture}(a). 
    There are user-side and item-side pre-training task in this stage. Both of the two tasks have the same model structure. The embedding layer is shared between the two tasks. The pre-training stage is based on a self-supervised contrastive learning strategy, which will be described in detail in subsection \ref{subsection:pre_training}. 

    \item \textbf{Fine-Tune Stage}, as shown in Figure \ref{fig:architecture}(b). 
    The model structure in fine-tune stage utilizes a dual double-tower framework to generate user and item representation vectors separately. The fine-tune and online serving operations will be illustrated in subsection \ref{subsection:fine_tune} and \ref{subsection:online_serving}. The embedding layer and network weights in fine-tune stage are restored from the pre-training stage, so that the model in fine-tune stage can reuse well-trained information from the entire sample space.
\end{itemize}

Both of the two stages utilize a \textbf{M}ulti-\textbf{L}ayer \textbf{S}cenario \textbf{A}daptive \textbf{T}ransfer Module (\textbf{ML-SAT}) to regulate the fine-grained and decoupled information transfer from whole scenarios to specific scenario. We minutely describe it in subsection \ref{subsection:ML-SAT}.

\subsection{Self-Supervised Framework Based on Contrastive Learning\label{subsection:pre_training}}

We introduce the paradigm of contrastive learning for our pre-training task and creatively correlate the multi-scenario matching framework and contrastive learning framework in a unified way. Training samples in pre-training stage are drawn from both labeled(clicked) and unlabeled(exposed but not clicked) data space.

As shown in Figure \ref{fig:architecture}(a), in the user side, behaviors of the same user in different scenario can be treated as a mechanism of data augmentation. After that, a Multi-Layer Scenario Adaptive Transfer Module (ML-SAT) is proposed to generate distinguishing representation vectors for the same user in different scenarios. Finally, the contrastive loss function\cite{chen2020simple} is introduced as the optimization loss to maximize agreements between different latent representations of same user in different scenarios. If a user $u$ accessed $k$ scenarios, we simply split these scenarios and make a combination of two to generate $C_k^2=\frac{k*(k-1)}{2}$ samples for contrastive learning.

In the item side, exposures and interactions of the same item in different scenario can also be treated as a mechanism of data augmentation. Siamese model network and contrastive loss function will be adopted to the pre-training process for the same item. Similar operations of sample combination are conducted to generate multiple samples for item side, just as user side did.

\subsubsection{Feature Composition and Embedding Layer}


In the user side, every training sample contains user profiles, scenario context features and two groups of user-scenario cross behavior features which are drawn from two scenarios separately. Specifically, user profiles contain age, gender, etc. Scenario context features mainly include scenario ID. User-scenario cross behavior features including user behavior sequences, category preferences and user statistic characteristics in the corresponding scenario. 

For user behavior sequence features, the list of feature field embeddings will be concatenated to form the item embedding for each item in the behavior sequence. The final user behavior sequence embeddings will be generated by sequence pooling strategies. We utilize self-attention mechanism\cite{devlin2018bert} as our sequence pooling operation, other strategies\cite{shen2021sar,lv2019sdm} can be investigated for better performance, which we remain for future research.

In the item side, item profiles contain item ID, item category ID, account ID, etc. item-scenario cross features mainly include item statistic characteristics in the corresponding scenario.

\begin{figure*}[tb]
  \setlength{\abovecaptionskip}{5pt}
  \setlength{\belowcaptionskip}{0pt}
  \centering
  \includegraphics[width=\linewidth]{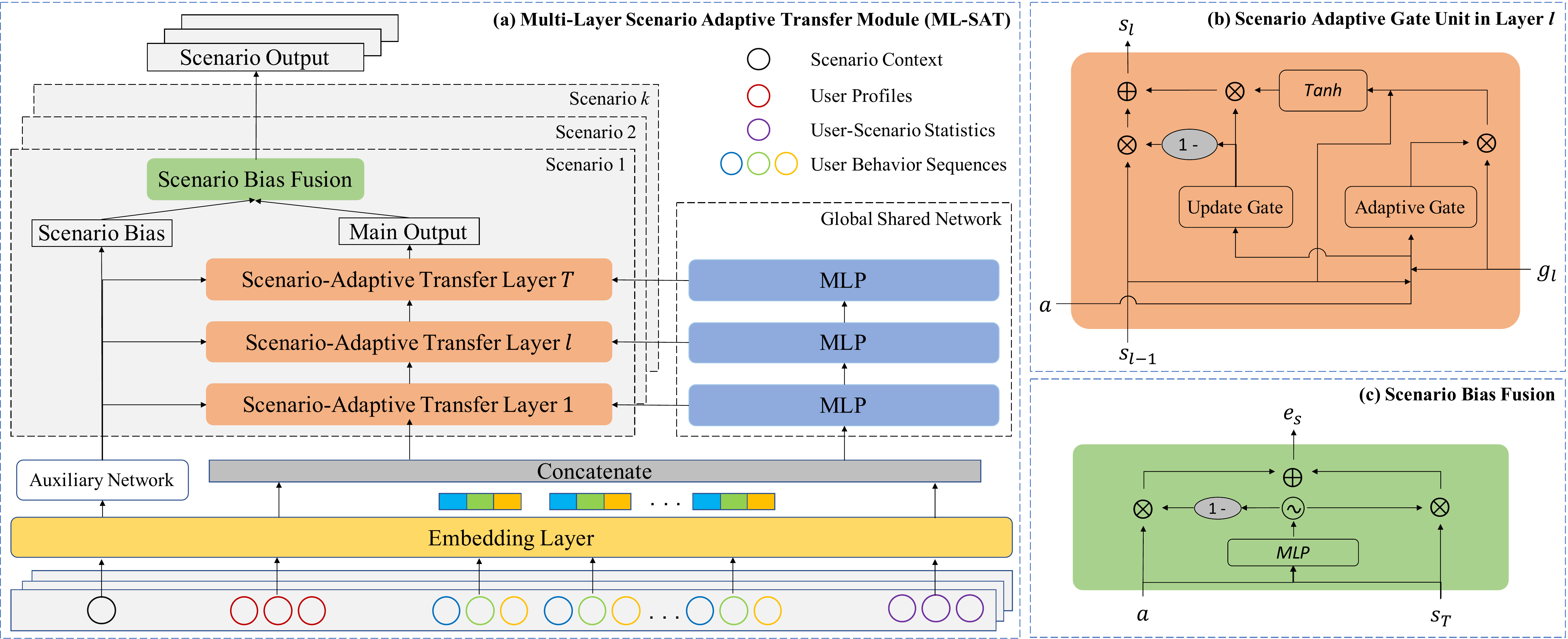}
  \caption{Overview of Multi-Layer Scenario Adaptive Transfer Module. We depict the model structure in user side. Item side's structure is identical with this. We describe the framework in one scenario in detail. Other scenarios will follow the same paradigm to generate corresponding vectors. (a) Multi-Layer Scenario Adaptive Transfer module. (b) Scenario-Adaptive Gate Unit in $l$th layer of ML-SAT. (c) Scenario bias fusion mechanism}
  \vspace{-10pt}
  \label{fig:ML_SAT}
\end{figure*}

To highlight the significance of scenario context features, we introduce a separate auxiliary network to model scenario characteristic, as formulated below:
\begin{equation}
    \setlength{\abovedisplayskip}{3pt}
    \setlength{\belowdisplayskip}{3pt}
    a = f(W_a x_a + b_a)
\end{equation}
where $x_a$ is the embedding of scenario context features, $f(*)$ is a multi-layer perceptron.



\subsubsection{Multi-Layer Scenario Adaptive Transfer Module (ML-SAT)\label{subsection:ML-SAT}}

We propose a novel network to extract representation vectors in corresponding scenario both for users and items. The structures between user side and item side are nearly the same, as shown in Figure \ref{fig:architecture}. Thus, we only describe the user side network in the following of the paper.

To make full use of whole-scenario information and regulate the fine-grained information transfer from whole scenarios to specific scenario, we introduce a \emph{global shared network} to learn the information of all scenarios, and propose a \textbf{M}ulti-\textbf{L}ayer \textbf{S}cenario \textbf{A}daptive \textbf{T}ransfer Module(\textbf{ML-SAT}) as the \emph{scenario specific network} for each individual scenario. As shown in Figure \ref{fig:ML_SAT}(a), the \emph{global shared network} is a multi-layer perceptrons shared by all the scenarios. Training samples coming from the whole scenarios will be fed into \emph{global shared network} to train a whole-scenario model. ML-SAT, which is parameter-specific in separate scenario, is trained with training samples coming from corresponding scenario. 

Motivated by GRU\cite{cho2014learning}, in each network layer, we design a \emph{scenario-adaptive gate unit} with explicit gate mechanisms to regulate the fine-grained information transfer from whole scenarios to specific scenario. As shown in Figure \ref{fig:ML_SAT}(b), the \emph{scenario-adaptive gate unit} can be formulated as follows:
\begin{equation}
    \setlength{\abovedisplayskip}{3pt}
    \setlength{\belowdisplayskip}{3pt}
    r_{l} = \sigma(W_{r}^{l}[g_{l}, s_{l-1}] + W_{br}a)
    \label{equation:rl}
\end{equation}
\begin{equation}
    \setlength{\abovedisplayskip}{3pt}
    \setlength{\belowdisplayskip}{3pt}
    h_{l} = tanh(W_{h}^{l}[r_{l} \cdot g_{l}, s_{l-1}])
    \label{equation:hl}
\end{equation}
\begin{equation}
    \setlength{\abovedisplayskip}{3pt}
    \setlength{\belowdisplayskip}{3pt}
    z_{l} = \sigma(W_{z}^{l}[g_{l}, s_{l-1}] + W_{bz}a) \label{equation:zl}
\end{equation}
\begin{equation}
    \setlength{\abovedisplayskip}{3pt}
    \setlength{\belowdisplayskip}{3pt}
    s_{l} = (1 - z_{l}) \cdot s_{l-1} + z_{l} \cdot h_{l}
    \label{equation:zl_1}
\end{equation}
where $g_{l}$ denote the $l$th layer output of \emph{global shared network}. $s_{l-1}$ denote the $(l-1)$th layer output of \emph{scenario specific network}. $W_{z}^{l}$ and $W_{bz}$ are the projection matrix weights and bias matrix weights of \emph{update gate} $z_{l}$ in $l$th layer. $W_{r}^{l}$ and $W_{br}$ are the projection matrix weights and bias matrix weights of \emph{adaptive gate} $r_{l}$ in $l$th layer. $a$ is the output of the scenario auxiliary network, which is introduced as a strong scenario indicator bias. 

In Equation (\ref{equation:rl}), \emph{adaptive gate} $r_{l}$ decides the degree of useful information the \emph{global shared network} can transfer. $h_{l}$ reflects the new adaptive state transfer information, considering the correlation between $g_{l}$ and $s_{l-1}$, as depicted in Equation (\ref{equation:hl}). With the \emph{update gate} $z_{l}$ as fusion weights, the new output $s_{l}$ can be updated by the fusion of $s_{l-1}$ and adaptive transfer output $h_{l}$. The network layer with \emph{scenario-adaptive gate unit} can be stacked to multiple layers for more progressive layered extraction\cite{tang2020progressive}.

\subsubsection{Scenario Bias Fusion}

After the multi-layer network with \emph{scenario-adaptive gate unit}, we can get the scenario main output of the corresponding scenario. Considering the significant importance of scenario context features in distinguishing scenario characteristics, we treat the output of scenario auxiliary network as scenario bias and fuse it with scenario main output to generate the final scenario-specific representation vector, as shown in Figure \ref{fig:ML_SAT}(c).
\begin{equation}
    \setlength{\abovedisplayskip}{2pt}
    \setlength{\belowdisplayskip}{1pt}
    e_{s} = \alpha \cdot s_{T} + (1 - \alpha) \cdot a
\end{equation}
\begin{equation}
    \setlength{\abovedisplayskip}{1pt}
    \setlength{\belowdisplayskip}{3pt}
    \alpha = \sigma(W_{o}[s_{T}, a])
    \vspace{-3pt}
\end{equation}
where $e_{s}$ is the final output, $s_{T}$ is the scenario main output, with $T$ as the final layer number. $a$ is the output of the scenario auxiliary network. $\sigma$ is the sigmoid function.

\subsubsection{Self-supervised Optimization Objective}

For each training sample with two groups of scenario-specific features in corresponding scenario, we can get two output vectors for the same user or item separately, denoted as $e_{s}^{i}$ and $e_{s}^{j}$. The objective of pre-training task is to extract agreements and model corrections between different scenarios. Thus, we adopt the same self-supervised contrastive loss of \cite{chen2020simple} as our optimizing loss. Specifically, for a minibatch training samples with batch size $N$, we can get $2N$ vectors after ML-SAT. we treat $e_{s}^{i}$ and $e_{s}^{j}$ as a positive pair. The other $2(N-1)$ scenario vectors are regarded as negative vectors. The loss function for a positive pair $(e_{s}^{i}, e_{s}^{j})$ is defined as
\begin{equation}
    \setlength{\abovedisplayskip}{3pt}
    \setlength{\belowdisplayskip}{3pt}
    \mathcal{L}_{ij} = -log\frac{exp(sim(e_{s}^{i}, e_{s}^{j})/\tau)}{\sum_{k=1, k \ne i}^{2N}exp(sim(e_{s}^{i}, e_{s}^{k})/\tau)}
\end{equation}
where $sim(e_{s}^{i}, e_{s}^{j}) = \frac{(e_{s}^{i})^T e_{s}^{j}}{\lVert e_{s}^{i} \rVert \lVert e_{s}^{j} \rVert}$. $\tau$ is the temperature parameter. The final loss is the sum of all losses in the minibatch, as denoted in equation (\ref{equation:final_loss}).
\begin{equation}
    \setlength{\abovedisplayskip}{3pt}
    \setlength{\belowdisplayskip}{3pt}
    \mathcal{L} = \sum_{k=1}^{N} (\mathcal{L}_{ij})^{(k)}
    \label{equation:final_loss}
\end{equation}

\subsection{Fine-Tune Stage\label{subsection:fine_tune}}

Most model components in fine-tune stage are identical with pre-training stage, such as ML-SAT module and Scenario Bias Fusion module. For brevity, we only describe components different from pre-training stage in the following subsections.

\subsubsection{Feature Composition and Embedding Layer}

\begin{figure*}[tb]
  \setlength{\abovecaptionskip}{5pt}
  \setlength{\belowcaptionskip}{3pt}
  \centering
  \includegraphics[width=0.96\linewidth]{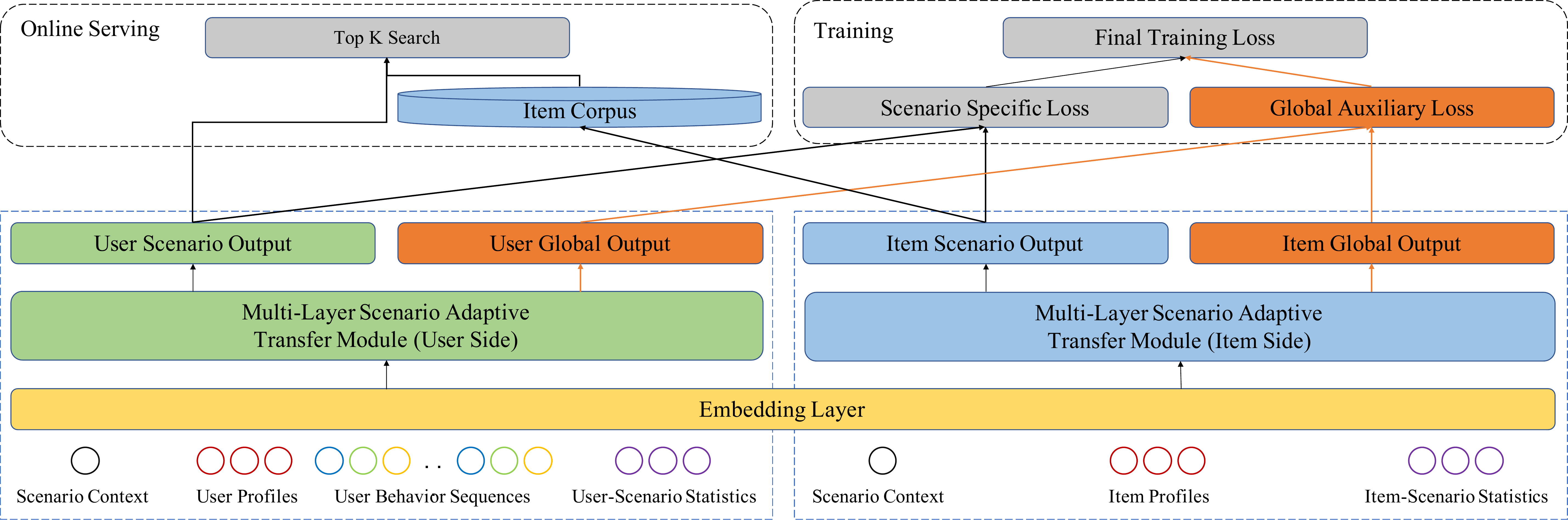}
  \caption{Fine-tune stage of SASS. Every user and item will output two vectors respectively, with the scenario specific loss and global auxiliary loss as optimization objectives. For online serving, we only employ scenario output of users and items.}
  \vspace{-10pt}
  \label{fig:fine_tune_task}
\end{figure*}

The training samples in fine-tune stage are labeled data drawn from target scenario. A user can only access one scenario at a time, so the features in one training sample only contain features in a single scenario. User side features contain user profiles, scenario context feature (scenario ID), user's behavior sequences, statistic features and prefer features of user in target scenario. Item side features contain item profiles, scenario context feature (scenario ID), statistic features and prefer features of item in target scenario. Then, as shown in Figure \ref{fig:fine_tune_task}, user side features and item side features are separately fed into embedding layer and upper ML-SAT modules to generate final scenario-specific vectors in corresponding scenario both for users and items. It should be emphasized that the embedding layer and networks weights of ML-SAT in fine-tune stage are restored from pre-training stage.


\subsubsection{Fine-Tune Optimization Objective} The final optimization loss in fine-tune stage is the combination of \emph{Scenario Specific Loss} and \emph{Global Auxiliary Loss}.

\emph{Scenario Specific Loss Function}: Similar with other matching tasks\cite{huang2020embedding,nigam2019semantic,zhang2020towards}, we adapt pairwise loss to optimize our fine-tune matching task. For a scenario $s$, the $k$th training sample in fine-tune stage is a triplet ($u_s^{k}$, $p_s^{k}$, $n_s^{k}$). $u_s^{k}$ denotes user representation vector, $p_s^{k}$ and $n_s^{k}$ denote corresponding positive and negative item vector, respectively. Negative items are randomly sampled from whole candidate set with negative sampling strategy\cite{mikolov2013efficient}. The scenario-specific loss function is defined as
\begin{equation}
    \setlength{\abovedisplayskip}{3pt}
    \setlength{\belowdisplayskip}{3pt}
    \mathcal{L}_{scenario} = \frac{1}{N} \sum_{k=1}^{N} log(1 + \sigma(sim(u_s^{k}, n_s^{k}) - sim(u_s^{k}, p_s^{k})))
\end{equation}
where $\sigma$ is sigmoid function and $sim(*)$ is cosine function.

\emph{Global Auxiliary Loss Function}: In fine-tune stage, the \emph{global shared network} will also output a representation vector $g_{T}$ for each sample, as shown in Figure \ref{fig:fine_tune_task}. The \emph{global shared network} is trained with samples from all scenarios. So the output $g_{T}$ can be treated as the user or item representations in a global perspective. 
Modeling the similarity of users and items in the global scope is beneficial to the convergence of training and performance improvement. So we introduce a global auxiliary loss defined as:
\begin{equation}
    \setlength{\abovedisplayskip}{3pt}
    \setlength{\belowdisplayskip}{3pt}
    \mathcal{L}_{auxiliary} = \frac{1}{N} \sum_{k=1}^{N} log(1 + \sigma(sim(u_g^{k}, n_g^{k}) - sim(u_g^{k}, p_g^{k})))
\end{equation}
where $u_g^{k}$, $p_g^{k}$, $n_g^{k}$ denote the representation vectors of a triplet (user, positive item, negative item) respectively, which are all generated though \emph{global shared network}. Finally, the loss function in fine-tune stage is formulated as below with a hyper-parameter $\beta$:
\begin{equation}
  \setlength{\abovedisplayskip}{3pt}
  \setlength{\belowdisplayskip}{0pt}
  \mathcal{L} = \mathcal{L}_{scenario} + \beta \cdot \mathcal{L}_{auxiliary}
\end{equation}

\subsection{Online Serving\label{subsection:online_serving}}

When SASS is trained, the whole model in fine-tune stage can be deployed for online serving. For a specific scenario $s$, all the items with their features will be fed into the model and generate item vectors $e_{s}$ from corresponding \emph{scenario specific network} in the item side architecture. Then, all the item vectors are saved as an item corpus. During online serving, when a user accesses scenario $s$, the user features are fed into SASS and the user vectors $u_{s}$ is generated from corresponding \emph{scenario specific network} in the user side architecture. Finally, an online real-time top-$k$ retrieval operation based on approximate near neighbor algorithms\cite{johnson2019billion} is implemented. These retrieved results are treated as candidates for subsequent ranking tasks.


\section{Experiments}
To adequately evaluate the proposed SASS model, we conduct experiments to answer the following research questions:
\begin{itemize} [topsep=2pt, leftmargin=8pt]
    \item How about SASS model compared with state-of-art matching models trained with one scenario data for each scenario or trained with whole-scenario data for all scenarios?
    \item How about SASS model compared with other state-of-art multi-scenario matching models?
    \item How about the impact of each part on the overall model?
\end{itemize}
\subsection{Experimental Settings}
\subsubsection{Datasets} 
We conduct experiments on our industrial dataset and two public datasets. Table \ref{table:datasets} summarizes the basic information of these datasets.

\begin{itemize} [topsep=2pt, leftmargin=8pt]
    \item \textbf{Ali-MSSV}. Our industrial Muiti-Scenario Short Video (MSSV) dataset in Taobao. Data from 2022-03-24 to 2022-04-04 is utilized for training and 2022-04-05 for testing. We evaluate models on two dense scenarios (denoted as \#A1 and \#A2) with abundant user behaviors and two spare scenarios (\#A3 and \#A4) with sparse data.
    \item \textbf{Ali-CCP}\cite{ma2018entire}. A public dataset released by Taobao with prepared training and testing set. We split the dataset into 3 scenarios according to \emph{scenario id}, denoted as \#B1 to \#B3 for simplicity.
    \item \textbf{Ali-Mama}\cite{gai2017learning}. A public dataset released by Alimama, an online advertising platform in China. Data from 2017-05-06 to 2017-05-011 is utilized for training and 2022-05-11 for testing. We arrange the dataset into 5 scenarios according to the \emph{city level}, denoted as \#C1 to \#C5 for simplicity.
\end{itemize}

\subsubsection{Competitors} 

We release two types of SASS models to compare with other methods. 

\begin{itemize} [topsep=2pt, leftmargin=8pt]
    \item \textbf{SASS-Base}: It is the model trained with labeled data without pre-training. We will evaluate the performance of SASS-Base on all of the three datasets. 
    \item \textbf{SASS}: It is our proposed model with two stages. The Ali-CCP dataset is unsuitable for pre-training task due to the lack of indispensable features, so we only evaluate the performance of SASS on our Ali-MSSV dataset and Ali-Mama dataset.
\end{itemize}

The compared \emph{Single-Scenario} matching models (trained with single-scenario data) are listed as follows:

\begin{itemize} [topsep=2pt, leftmargin=8pt]
    \item \textbf{YoutubeDNN}: YoutubeDNN\cite{davidson2010youtube} adopts average pooling to extract user’s interest with a sampled softmax loss to optimize similarities between users and items. 
    \item \textbf{DSSM}: DSSM\cite{huang2013learning} builds a relevance score model to extract user and item representations with double-tower architecture. 
    \item \textbf{BST}: BST\cite{chen2019behavior} leverages transformer to build the user behavior sequence. In this paper we use the inner product of user and item representations instead of MLP.
    \item \textbf{MIND}: MIND\cite{li2019multi} clusters users' multiple interests by capsule network to improve the effect of multi-interest promotion.
\end{itemize}

The model ment above will also be trained with all-scenario data for \emph{Mix-Scenario} versions.
\begin{table}[h]
  \setlength{\abovecaptionskip}{1pt}
  \setlength{\belowcaptionskip}{0pt}
  \centering
  \caption{Basic information of three datasets}
  \setlength{\tabcolsep}{1.4mm}{
    \begin{tabular}{ccccc}
        \toprule
              & Scenario & User & item  & Samples \\
        \midrule
        \multirow{4}[2]{*}{Ali-MSSV (\#A)} 
            & \#A1  & 10.1M & 2.5M  & 630M \\
            & \#A2  & 30.1M  & 5.6M   & 1.2B \\
            & \#A3  & 2.1M  & 0.53M   & 100.2M \\
            & \#A4  & 1.5M  & 0.39M   & 100M \\
        \midrule
        \multirow{3}[2]{*}{Ali-CCP  (\#B)} 
            & \#B1  & 0.13M & 1.98M & 0.63M \\
            & \#B2  & 0.18M & 2.44M & 1M \\
            & \#B3  & 50k   & 0.2M  & 13k \\
        \midrule
        \multirow{5}[2]{*}{Ali-Mama (\#C)} 
            & \#C1  & 40K   & 50k   & 94k \\
            & \#C2  & 0.15M & 0.13M & 0.42M \\
            & \#C3  & 80K   & 89k   & 0.23M \\
            & \#C4  & 65K   & 77k   & 0.19M \\
            & \#C5  & 0.14M & 0.12M & 0.35M \\
        \bottomrule
    \end{tabular}
  }
    \vspace{-11pt}
  \label{table:datasets}
\end{table}

The multi-scenario matching models are listed as follows. To the best of our knowledge, most of the existing multi-scenario models are mainly proposed for ranking problems, Therefore, for original multi-scenario ranking modes, some necessary but slight modifications are made for unified evaluation on matching tasks, denoted by a postfix \textbf{-M} for each model.
\begin{itemize} [topsep=2pt, leftmargin=8pt]
    \item \textbf{SAR-Net-M}: Modified version of SAR-Net\cite{shen2021sar} which propose a multi-scenario architecture for scenario information migration with scenario-specific users' behaviors and attention mechanism;
    \item \textbf{STAR-M}: Modified version of STAR\cite{sheng2021one} which constructs star topology for multi scenarios and operate an element-wise operation to control information transferred from central network to specific networks;
    \item \textbf{HMoE-M}: Modified version of HMoE\cite{li2020improving} which constructs relationship between multiple scenarios in label implicit space through stacked model;
    \item \textbf{ZEUS-M}: Modified version of ZEUS\cite{gu2021self} which learns unlabeled data through a self-supervised \emph{Next-Query Prediction} task;
    \item \textbf{ICAN}: ICAN\cite{xie2020internal} is one of the SOTA models for multi-domain matching, which is most related works of our task. It highlights the interactions between feature fields in different domains for cold-start matching.
\end{itemize}

\subsubsection{Parameter Settings and Metrics} 

For all methods, the truncation length of user behavior is 50. AdamGrad is used as the optimizer with learning rate of 0.001 for all methods and the batch size is 512. Hit Ratio (HR) and Normalized Discounted Cumulative Gain (NDCG) are adopted as the performance metrics. We set HR@20 and NDCG@20 in top-20 matching results as final metrics.

\subsection{Overall Experimental Results: RQ1 \& RQ2}

\begin{table*}[tbp]
  \setlength{\abovecaptionskip}{3pt}
  \setlength{\belowcaptionskip}{3pt}
  \centering
  \caption{Performance of different models on three datasets. Single-scenario models are trained with data in individual scenario independently while mix-scenario models are trained with data from all scenarios. Multi-Scenario models are trained with multi-scenario data in unified frameworks. Due to the lack of indispensable features for pre-traing task on Ali-CCP dataset, we only evaluate the performance of SASS on Ali-MSSV and Ali-Mama datesets.}
  \resizebox{\linewidth}{!}{ 
        \begin{tabular}{cc|cccc|cccc|ccccccc}
        \toprule
            \multicolumn{2}{c|}{\multirow{2}[4]{*}{}} & \multicolumn{4}{c|}{single-scenario models} & \multicolumn{4}{c|}{mix-scenario models} & \multicolumn{7}{c}{multi-scenario models} \\
           \cmidrule{3-17} \multicolumn{2}{c|}{} & \small MIND & \small DSSM &  \small BST &  \small YoutubeDNN &  \small MIND &  \small DSSM &  \small BST &  \small YoutubeDNN &  \small SAR-Net-M &  \small STAR-M &  \small HMoE-M &  \small ZEUS-M &  \small ICAN &  \small SASS-Base &  \small \textbf{SASS} \\
        \midrule
        \multirow{2}[1]{*}{\#A1} & HR@20 & 0.052 & 0.039 & 0.042 & 0.034 & 0.044 & 0.032 & 0.035 & 0.027 & 0.079 & \underline{0.082} & 0.061 & 0.072 & 0.072 & 0.082  & \textbf{0.087} \\
              & NDCG@20 & 0.021 & 0.012 & 0.014 & 0.015 & 0.024 & 0.010 & 0.012 & 0.009 & 0.033 & \underline{0.036} & 0.022 & 0.031 & 0.025 & 0.039 & \textbf{0.042} \\
        \multirow{2}[0]{*}{\#A2} & HR@20 & 0.032 & 0.016 & 0.029 & 0.023 & 0.027 & 0.012 & 0.023 & 0.016 & \underline{0.035} & 0.034 & 0.027 & 0.032 & 0.031 & 0.034 & \textbf{0.043} \\
              & NDCG@20 & 0.014 & 0.009 & 0.013 & 0.012 & 0.011 & 0.007 & 0.008 & 0.004 & 0.014 & 0.012 & 0.011 & 0.014 & \underline{0.016} & 0.019 & \textbf{0.029} \\
        \multirow{2}[0]{*}{\#A3} & HR@20 & 0.041 & 0.032 & 0.036 & 0.029 & 0.045 & 0.033 & 0.041 & 0.028 & 0.042 & \underline{0.045} & 0.041 & 0.039 & 0.039 & 0.047 & \textbf{0.068} \\
              & NDCG@20 & \underline{0.023} & 0.017 & 0.021 & 0.015 & 0.021 & 0.017 & 0.019 & 0.013 & 0.021 & 0.019 & 0.017 & 0.016 & 0.013 & 0.025 & \textbf{0.037} \\
        \multirow{2}[1]{*}{\#A4} & HR@20 & 0.017 & 0.010 & 0.012 & 0.006 & 0.023 & 0.011 & 0.012 & 0.009 & 0.027 & 0.021 & 0.029 & \underline{0.031} & 0.025 & 0.032 & \textbf{0.047} \\
              & NDCG@20 & 0.007 & 0.003 & 0.007 & 0.002 & 0.011 & 0.007 & 0.004 & 0.005 & 0.012 & \underline{0.013} & 0.010 & 0.012 & 0.010 & 0.017 & \textbf{0.023} \\
        \midrule
        \multirow{2}[1]{*}{\#B1} & HR@20 & 0.153 & 0.112 & 0.144 & 0.120 & 0.164 & 0.133 & 0.156 & 0.137 & 0.185 & 0.211 & 0.194 & \underline{0.212} & 0.175 & \textbf{0.234} & \textbf{-} \\
              & NDCG@20 & 0.061 & 0.043 & 0.057 & 0.049 & 0.091 & 0.052 & 0.069 & 0.062 & 0.082 & 0.102 & 0.097 & \underline{0.117} & 0.083 & \textbf{0.137} & \textbf{-} \\
        \multirow{2}[0]{*}{\#B2} & HR@20 & 0.191 & 0.143 & 0.167 & 0.136 & 0.212 & 0.168 & 0.193 & 0.151 & \underline{0.241} & 0.227 & 0.219 & 0.231 & 0.199 & \textbf{0.252} & \textbf{-} \\
              & NDCG@20 & 0.121 & 0.093 & 0.103 & 0.092 & 0.137 & 0.112 & 0.117 & 0.096 & 0.117 & 0.106 & 0.094 & \underline{0.123} & 0.104 & \textbf{0.131} & \textbf{-} \\
        \multirow{2}[1]{*}{\#B3} & HR@20 & 0.043 & 0.029 & 0.037 & 0.021 & 0.067 & 0.042 & 0.058 & 0.039 & 0.074 & 0.069 & \underline{0.081} & 0.079 & 0.069 & \textbf{0.092} & \textbf{-} \\
              & NDCG@20 & 0.019 & 0.013 & 0.013 & 0.011 & 0.029 & 0.019 & 0.028 & 0.021 & 0.032 & 0.041 & \underline{0.043} & 0.037 & 0.033 & \textbf{0.051} & \textbf{-} \\
        \midrule
        \multirow{2}[1]{*}{\#C1} & HR@20 & 0.213 & 0.176 & 0.192 & 0.172 & 0.232 & 0.179 & 0.193 & 0.132 & 0.239 & \underline{0.242} & 0.219 & 0.221 & 0.203 & 0.243 & \textbf{0.269} \\
              & NDCG@20 & 0.117 & 0.098 & 0.114 & 0.079 & 0.125 & 0.093 & 0.106 & 0.056 & 0.119 & 0.107 & \underline{0.129} & 0.114 & 0.105 & 0.132 & \textbf{0.137} \\
        \multirow{2}[0]{*}{\#C2} & HR@20 & 0.179 & 0.127 & 0.142 & 0.114 & 0.155 & 0.107 & 0.123 & 0.081 & 0.147 & 0.173 & 0.183 & 0.192 & \underline{0.204} & 0.209 & \textbf{0.227} \\
              & NDCG@20 & 0.083 & 0.054 & 0.078 & 0.042 & 0.069 & 0.042 & 0.067 & 0.033 & 0.084 & 0.088 & 0.092 & 0.097 & \underline{0.099} & 0.107 & \textbf{0.119} \\
        \multirow{2}[0]{*}{\#C3} & HR@20 & 0.256 & 0.237 & 0.217 & 0.193 & 0.251 & 0.241 & 0.224 & 0.201 & 0.241 & 0.269 & 0.243 & 0.261 & \underline{0.271} & 0.274 & \textbf{0.289} \\
              & NDCG@20 & 0.131 & 0.119 & 0.095 & 0.088 & 0.127 & 0.119 & 0.107 & 0.081 & 0.132 & \underline{0.139} & 0.137 & 0.122 & 0.138 & 0.142 & \textbf{0.151} \\
        \multirow{2}[0]{*}{\#C4} & HR@20 & 0.225 & 0.193 & 0.201 & 0.165 & 0.231 & 0.201 & 0.215 & 0.177 & 0.239 & 0.241 & \underline{0.251} & 0.247 & 0.221 & 0.259 & \textbf{0.267} \\
              & NDCG@20 & 0.131 & 0.099 & 0.123 & 0.077 & 0.127 & 0.113 & 0.093 & 0.081 & 0.132 & 0.114 & 0.132 & 0.129 & \underline{0.142} & 0.134 & \textbf{0.142} \\
        \multirow{2}[1]{*}{\#C5} & HR@20 & 0.191 & 0.167 & 0.188 & 0.143 & 0.185 & 0.155 & 0.147 & 0.132 & 0.204 & 0.212 & 0.209 & 0.189 & \underline{0.217} & 0.227 & \textbf{0.241} \\
              & NDCG@20 & 0.082 & 0.077 & 0.112 & 0.059 & 0.104 & 0.082 & 0.069 & 0.055 & 0.093 & 0.112 & 0.103 & 0.104 & \underline{0.121} & 0.132 & \textbf{0.137} \\
              
        \bottomrule
        \end{tabular}
    }
    \vspace{-8pt}
  \label{table:method_results}
\end{table*}

\begin{table}[tb]
  \setlength{\abovecaptionskip}{1pt}
  \setlength{\belowcaptionskip}{1pt}
  \centering
  \caption{Ablation study of Scenario-Adaptive Gate Unit.}
    \setlength{\tabcolsep}{0.5mm}{
        \begin{tabular}{l|cc|cc}
            \toprule
                  & \multicolumn{2}{c|}{\#A1} & \multicolumn{2}{c}{\#A3} \\
                  & HR@20 & NDCG@20 & HR@20 & NDCG@20 \\
            \midrule
            SASS$^{*}$ & 0.054 & 0.021 & 0.029 & 0.019 \\
            SASS$^{*}$+Production Gate & 0.059 & 0.022 & 0.033 & 0.018 \\
            SASS$^{*}$+Simnet Gate & 0.065 & 0.027 & 0.037 & 0.017 \\
            SASS$^{*}$+Sigmoid Gate & 0.071 & 0.029 & 0.042 & 0.023 \\
            SASS-Base  & \textbf{0.082} & \textbf{0.039} & \textbf{0.047}  & \textbf{0.025}  \\
            \bottomrule
        \end{tabular}
    }
    \vspace{-15pt}
  \label{table:ablation_study_gate}%
\end{table}

We evaluate our methods with other compared models, As illustrated in Table \ref{table:method_results}. We can summarize three significant observations: (1) The experiment performances of mix-scenario models are generally better than single-scenario models in sparse scenarios (such as scenario \#A4, \#B1). We suspect that it is challenging for sparse scenarios to train perfect models with limited training samples. Additional samples coming from other scenarios can partly improve the performance of sparse scenarios. However, in dense scenarios (such as \#A1, \#C2), mix-scenario models perform worse than single-scenario models. One possible reason may be that the crude mixture of multi-scenario samples introduces non-negligible noise data, which is harmful to the model performances. These two observations are consistent with the conclusions in SAR-Net\cite{shen2021sar}. (2) Unified multi-scenario modes all get better performances than single-scenario models and mix-scenario models. The results indicate that the strategies with information share and modeling of scenario commonalities and distinctions are beneficial to solving multi-scenario problems. (3) SASS-Base (without pre-training) outperforms other compared models (single-scenario, mix-scenario and multi-scenario) in nearly all scenarios (with comparable performance with SAR-Net-M and STAR-M in scenario \#A2). It shows that SASS-Base has a leading performance over other multi-scenario models in building commonalities and characteristics of different scenarios. Moreover, SASS (with pre-traing) can further improve the performance, especially for sparse scenarios. The performance gain of SASS in scenarios \#A3 is much higher than that in other scenarios, indicating the excellent improvements of pre-training task in sparse scenarios.
 
\subsection{Ablation Study: RQ3}

\subsubsection{Scenario-Adaptive Gate Unit} 

\emph{Scenario-Adaptive Gate Unit} controls the information transfer from whole scenarios to specific scenario, which is essential for modeling scenario commonalities and distinctions. In this subsection, we investigate different transfer gate mechanisms and compare their performance on the SASS-Base model (SASS without pre-training stage). 
\begin{figure}[tb]
  \setlength{\abovecaptionskip}{0pt}
  \setlength{\belowcaptionskip}{0pt}
  \centering
  \includegraphics[width=0.95\linewidth]{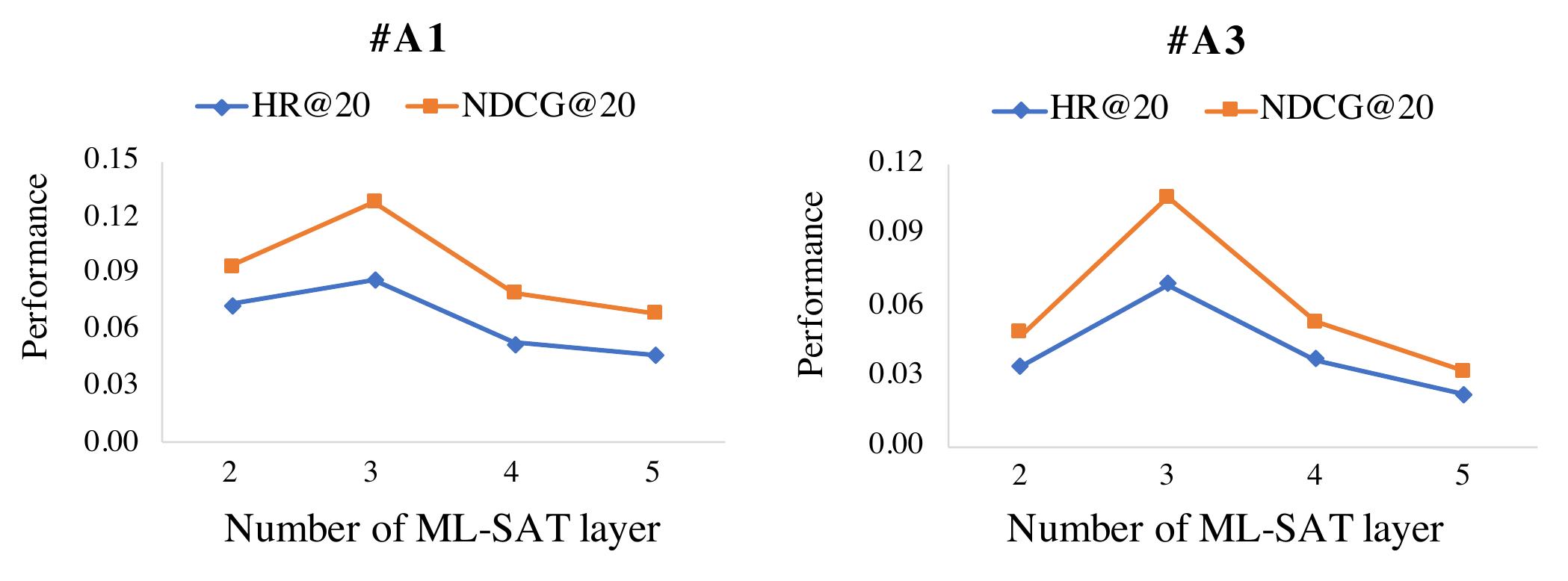}
  \caption{Ablation study of different layer number.}
  \label{fig:ablation_study_layer} 
\end{figure}
The baseline model is SASS$^{*}$, which is a SASS-Base version without transfer gate. SASS$^{*}$ can be treated as a variant framework of MOE\cite{shazeer2017outrageously}. In particular, we compare the following settings: (1) Sigmoid Gate: concatenating scenario specific layer's output $s_l$ with global shared layer's output $g_l$ and feeding the results into MLP with sigmoid to control information transfer. SASS$^{*}$+Sigmoid Gate can be treated as a variant framework of MMoE\cite{ma2018modeling} or Cross-Stitch\cite{misra2016cross}, which is also the fundamental transfer and fusion structure in SAR-Net\cite{shen2021sar} and HMoE\cite{li2020improving}. (2) Production Gate: taking the element-wise production of scenario specific layer's output $s_l$ and global shared layer's output $g_l$ to generate the final scenario output, which can be partly treated as an equivalent of information mapping of STAR\cite{sheng2021one}. (3) Simnet Gate: an updated version for the fusion of information with the concatenations of element-wise production, element-wise subtraction and element-wise addition as the input. Then the input is feed into MLP with sigmoid. It can be summarized from Table \ref{table:ablation_study_gate} that (1) SASS$^{*}$ performs worst and SASS$^{*}$ with other gate mechanisms are beneficial to scenario-specific modeling both in \#A1 and \#A3. (2) The proposed SASS-Base model achieves the best performance. It shows that the fine-grained and decoupled gate mechanism in ML-SAT can get better control on information transfer from whole scenarios to individual scenario.

\begin{table}[tb]
  \setlength{\abovecaptionskip}{3pt}
  \setlength{\belowcaptionskip}{0pt}
  \centering
  \caption{Ablation study of different pre-training strategies. \emph{Next Video Prediction} is the strategy introduced by ZEUS}
  \setlength{\tabcolsep}{0.6mm}{
    \begin{tabular}{c|cc|cc}
        \toprule
            & \multicolumn{2}{c|}{\#A1} & \multicolumn{2}{c}{\#A3} \\
            & HR@20    & NDCG@20  & HR@20    & NDCG@20 \\
        \midrule
            SASS-Base & 0.082 & 0.039 & 0.047 & 0.025 \\
            Next Video Prediction & 0.083 & 0.033 & 0.052 & 0.031 \\
            SASS  & \textbf{0.087} & \textbf{0.042} & \textbf{0.068} & \textbf{0.037} \\
        \bottomrule
    \end{tabular}
  }
  \vspace{-6pt}
  \label{table:ablation_study_pre_training}
\end{table}

\begin{table}[tb]
  \setlength{\abovecaptionskip}{1pt}
  \setlength{\belowcaptionskip}{1pt}
  \centering
  \caption{Ablation study of different item representation strategies}
  \setlength{\tabcolsep}{0.6mm}{
    \begin{tabular}{c|cc|cc}
        \toprule
            & \multicolumn{2}{c|}{\#A1} & \multicolumn{2}{c}{\#A3} \\
            & HR@20 & NDCG@20 & HR@20 & NDCG@20 \\
        \midrule
            Single Item Embedding & 0.081 & 0.037 & 0.054 & 0.033 \\
            SASS  & \textbf{0.087} & \textbf{0.042} & \textbf{0.068} & \textbf{0.037} \\
        \bottomrule
    \end{tabular}
  }
  \vspace{-13pt}
  \label{table:ablation_study_item}
\end{table}

\subsubsection{Different layer number of ML-SAT}

Motivated by PLE\cite{tang2020progressive}, the Scenario-Adaptive Transfer Layer can be stacked multiple layers for better performances. Thus, we set SASS-Base model with different layers for comparison. The results can be shown in Figure \ref{fig:ablation_study_layer}. The performances become better when the number of layers are stacked from 2 to 3, while getting worse with the further increase of layer number. We suspect that, with the layer number increasing, the distinctions of scenario-specific representation are decreasing if much more information is transferred from whole scenarios to specific scenario.

\subsubsection{Self-supervised Learning based Pre-training}

In this part, we investigate the performances of different self-supervised strategies in pre-training stage. We set SASS-Base (model without pre-training) as baseline model. We compare SASS-Base with SASS (with pre-training task) and another self-supervised \emph{Next Video Prediction} task, which is introduced in ZEUS\cite{gu2021self}. The results shown in Table \ref{table:ablation_study_pre_training} exhibit the superior performance of SASS, especially in spare scenarios (such as \#A3).

\subsubsection{Representations for items in different scenarios} 

To certify the performance of generating distinguishing item representations in different scenarios, we implement a variant of SASS with only one single item representation vector shared in multiple scenarios. Results in Table \ref{table:ablation_study_item} indicate that the performances gets better when generating different item representations in multiple scenarios. The reason is that when we consider the different representations of items in various scenarios, the characteristics of individual scenario can be more directly captured.

\subsubsection{Scenario Bias Fusion \& Global Auxiliary Loss Task} 

Scenario Bias Fusion is proposed to highlight the importance of scenario context information, while the global auxiliary loss task is introduced to strengthen the learning of global shared network. Both of the two strategies are expected to improve the whole performances of SASS. Experimental results in Table \ref{table:ablation_study_auxiliary} demonstrate our hypothesises.

\begin{table}[tb]
  \setlength{\abovecaptionskip}{4pt}
  \setlength{\belowcaptionskip}{0pt}
  \centering
  \caption{Ablation study of scenario bias fusion and global auxiliary loss task}
  \setlength{\tabcolsep}{0.3mm}{
    \begin{tabular}{c|cc|cc}
        \toprule
            & \multicolumn{2}{c|}{\#A1} & \multicolumn{2}{c}{\#A3} \\
            & HR@20 & NDCG@20 & HR@20 & NDCG@20 \\
        \midrule
            No Global Auxiliary Task & 0.045 & 0.018 & 0.036 & 0.021 \\
            No Scenario Bias Fusion & 0.075 & 0.031 & 0.039 & 0.024 \\
            SASS  & \textbf{0.087} & \textbf{0.042} & \textbf{0.068} & \textbf{0.037} \\
        \bottomrule
    \end{tabular}
  }
  \vspace{-4pt}
  \label{table:ablation_study_auxiliary}
\end{table}

\subsection{Online Deployment Test} 

Since August 2021, we have conducted online A/B tests and successfully developed SASS model on Taobao in Alibaba, which contains multiple short video recommendation scenarios. We collect the overall improvements of A/B tests in each industrial scenarios, where the base model is the double-tower matching model\cite{huang2020embedding}. The online evaluation metric is AWT (average watching time per user) and CTR (the number of clicks over the number of video impressions). As shown in Table \ref{table:ablation_study_online}, the online results have demonstrated the feasibilities and effectiveness of our proposed SASS model on real industrial recommendation systems.

\begin{table}[tb]
  \setlength{\abovecaptionskip}{0pt}
  \setlength{\belowcaptionskip}{0pt}
  \centering
  \caption{AWT and CTR gains in online short video recommendation platform of Taobao, Alibaba}

  \resizebox{\linewidth}{!}{ 
    \begin{tabular}{c|cc|cc|cc|cc}
        \toprule
        Scenarios & \multicolumn{2}{c|}{\#A1} & \multicolumn{2}{c|}{\#A2} & \multicolumn{2}{c|}{\#A3} & \multicolumn{2}{c}{\#A4} \\
        \midrule
        Metrics & AWT & CTR & AWT & CTR & AWT & CTR & AWT & CTR \\
        Gains  & +8.3\% & +16.3\% & +4.5\% & +5.3\% & +3.1\% & +1.2\% & +15.2\% & +16.2\% \\
        \bottomrule
    \end{tabular}
  }
  \vspace{-10pt}
  \label{table:ablation_study_online}%
\end{table}%

\section{Conclusion}

In this paper, we propose the Scenario-Adaptive and Self-Supervised (SASS) model to tackle three core problems of multi-scenario modeling mentioned above. To model multi-scenario commonalities and distinctions, SASS build a global shared network for all scenarios and a Multi-Layer Scenario Adaptive Transfer Module (ML-SAT) as scenario-specific network for each scenario. In ML-SAT, the Scenario-Adaptive Gate Unit is introduced to select and control information transfer from global shared network to scenario-specific network in a much fine-grained and decoupled way. To sufficiently exploit the power of entire space samples, a two-stage training framework including pre-training and fine-tune is introduced. The pre-training stage is based on a scenario-supervised contrastive learning task with the training samples drawn from labeled and unlabeled data spaces. Moreover, the model architecture of SASS is created symmetrically both in user side and item side, so that we can get distinguishing representations for items in individual scenarios.
The experimental results on both offline datasets (industrial and public) and online A/B tests demonstrate the superiority of SASS over state-of-the-art methods for solving multi-scenario problems. SASS has been deployed on the online short video recommendation platform in Taobao, bringing more than 8\% improvement on AWT.

\bibliography{acmart}
\end{document}